\documentclass[11pt]{article}
\renewcommand{\baselinestretch}{1.3}
\def\be{\begin{equation}}
\def\ee{\end{equation}}
\def\ba{\begin{eqnarray}}
\def\ea{\end{eqnarray}}

\def\nn{\nonumber}
\input{epsf}
\begin{document}
\title{Quasi-additivity of Tsallis entropies and correlated subsystems}
\author{S. Asgarani \thanks{email: sasgarani@ph.iut.ac.ir} ,
       \ \ B. Mirza \thanks{email: b.mirza@cc.iut.ac.ir} \\ \\
{\it Department of  Physics, Isfahan University of Technology}\\
{\it Isfahan 84156-83111,  Iran } \\}

\date{}
\vspace{4cm}
\maketitle{$\hspace{6.8cm}$\Large{Abstract}}\\
 We use Beck's quasi-additivity of Tsallis
entropies for $n$ independent subsystems to show that like the
case of $n=2$, the entropic index $q$ approaches 1 by increasing
system size. Then, we will generalize that concept to correlated
subsystems to find that in the case of correlated subsystems,
when system size increases, $q$ also approaches a value
corresponding to the additive case.
\vspace{.5cm}

\vspace{1.5cm}
\section{Introduction}
The formalism of non-extensive statistical mechanics has been
developed over the past two decades as a beautiful generalization
of ordinary statistical mechanics \cite{2,3,4}. It is based on
the extremization of the Tsallis entropies
\be\label{11}\\
S_{q}=\frac{1-\sum_{i=1}^Wp_{i}^q}{q-1}~,\\
\ee subject to suitable constraints. Here, the $p_i$ are the
probabilities of physical microstates, and q is the entropic
index. The Tsallis entropies reduce to the Boltzmann-Gibbs (or
Shannon) entropy $S_{BG}=\sum_{i=1}^Wp_{i}\ln{p_{i}}$ for
$q\rightarrow1$. Growing experimental evidence indicates that
$q\neq1$ yields a correct description of many complex physical
phenomena \cite{5}. Tsallis' original suggestion was that this
approach might be relevant for equilibrium systems with long-range
interactions, but recently it was also pointed out that the
formalism comes into play when systems are far from equilibrium.
For example, for systems with fluctuating mean free path
\cite{6,7} and temperature or energy dissipation rate \cite{8,9},
Tsallis entropy may be relevant. The common feature in the
 examples of non-equilibrium systems mentioned is that the
parameter $q$ can be expressed by the relative variance of the
fluctuations of a parameter $X$
\begin{equation}\label{12}
q=1\pm\frac{\langle{X^2}\rangle-{\langle{X}\rangle}^2}{{\langle{X}\rangle}^2}~,
\end{equation}
provided that $X$ is ${\chi}^2$ (or gamma) distributed. Here the
signs $`+$' $(`-$') refer to $q>1$ ($q<1$) cases, respectively.
From this point of view, the parameter $q$ can be treated in such
systems as a measure of fluctuations of a parameter $X$ \cite{6,7}.\\
Recently Beck \cite{10} has suggested that for systems with
fluctuations in temperature or energy dissipation rate, it is
possible to make the Tsallis entropies quasi-additive by choosing
different entropic indices at different spatial scales:
\begin{equation}\label{13}
 S_{q}^A+S_{q}^B=S_{q'}^{A+B}~,
\end{equation}
where q is used for statistically independent identical
subsystems $A$ and $B$ and $q'$ for a composed system $A+B$. Eq.
(\ref{13}) for $q$ close to 1 results in
\begin{equation}\label{14}
q'-1=(q-1)\frac{\langle{B_i^2}\rangle}{\langle{B_i^2}\rangle+{\langle{B_i}\rangle}^2}~,
\end{equation}
where $B_i:=\ln{p_i}$ is the so called "bit number" \cite{11}, its
negative expectation $-<B_i>:=-\sum_ip_i\ln{p_i}$ is the Shannon
entropy, and its variance $-<B_i^2>:=-\sum_ip_i(\ln{p_i})^2$ is
related to fluctuations in entropy. From Eq.~(\ref{14}), it is
clear that if $q>1$, then $q>q'>1$. This is what Beck interpreted
as scale dependence. In other words, $q'$ related to the composed
system is less than $q$ for each subsystem. Moreover, there are
experimental examples showing that $q$ is monotonously decreasing
as a function of distance $r$ \cite{12}. This idea has been used
in \cite{16} for independent \textit{non-identical} subsystems
with different entropic indices $q_1$ and $q_2$, where a power~law
for entropic index was extended as a function of distance $r$. The
Tsallis entropies are non-extensive (there is little difference
between extensivity and additivity and in this paper we will use
extensivity and additivity interchangably). Given two independent
subsystems $A$ and $B$ with probabilities $p_i^A$ and $p_i^B$,
respectively, the entropy of the composed system $A+B$ (with
probabilities $p_{ij}^{A+B}=p_i^Ap_i^B$) satisfies
\be\label{15}\\
S_{q}^{A+B}=S_{q}^{A}+S_{q}^{B}+(1-q)S_{q}^{A}S_{q}^{B}~.\ee
Hence, for independent subsystems, there is additivity only for
$q=1$. More recently, another interesting concept has been
reported \cite{13,14,15}. Within this approach, equal and
distinguishable subsystems can be strongly (globally) correlated
such that for an adequate value of $q\neq1$, $S_q$ becomes
strictly additive. This means that additivity can exist even for
correlated subsystems.

There are three demands to be considered: independency of
subsystems, the same $q$ for the composed system and subsystems,
and additivity of $S_q$. These demands can not coexist except when
$q\rightarrow1$ (Boltzmann-Gibbs entropy). Most early papers on
non-extensive statistical mechanics deal with the first and the
second demands and give~up the third, Beck \cite{10} keeps the
first and the third, while Tsallis \cite{13} considers the second
and the third.\\

This paper is organized as follows. In section 2, Beck's work is
generalized by including $n$ subsystems instead of the original
two subsystems. In section 3, we will attempt to use
quasi-additivity property (Eq. (\ref{13})) for correlated
subsystems where $q$ is assumed to be close to the additive case
$q^*$. In section 4, a method of finding probabilities satisfying
additivity relation is presented. In section 5, some numerical
solutions of quasi-additivity relation for two correlated equal
binary subsystems are given and, finally, we have a conclusion in
section 6.

\section{Quasi-additivity for n independent identical subsystems}
\begin{figure}
\centering \epsfysize=5cm\epsfysize=5cm\epsffile{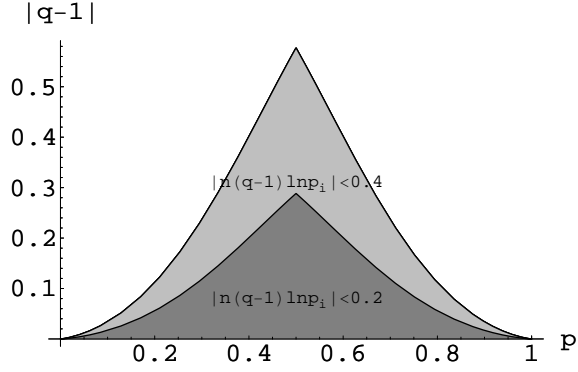}
\caption{\footnotesize Permissible region for $|q-1|$ plotted and
colored as a function of p, for two-state subsystems with the
probabilities $p$ and $1-p$. }\label{**}
\end{figure}

Quasi-additivity relation (Eq. (\ref{13})) for $n$ identical
subsystems can be written as:
\begin{equation}\label{21}
\sum_{r=1}^nS_q^{A_r}=S_{q'}^{A_1+A_2+\ldots+A_n}~,
\end{equation}
where $S_q^{A_r}$ is the entropy of one of the subsystems and
$S_{q'}^{A_1+\ldots+A_n}$ is the entropy of the composed system.
Our assumption is that $q$ and $q'$ are different and are close to
1. Parallel to what Beck did \cite{10}, one can obtain
\begin{equation}\label{22}
\sum_ip_i^q=\sum_ip_i\exp[{(q-1)\ln{p_i}}]=1+(q-1)\sum_ip_i\ln{p_i}+\frac{1}{2}{(q-1)}^2\sum_ip_i{(\ln{p_i})}^2+\ldots~,
\end{equation}
where the dots denote higher-order terms in $q-1$. We can neglect
the higher-order terms only if
\begin{equation}\label{22*}
|(q-1)\ln{p_i}|<<1 ,\hspace{.5cm}\forall\hspace{.1cm}{i}
\end{equation}
or, if alternatively,
\begin{equation}\label{22**}
p_i>>\exp\frac{-1}{|q-1|}~.
\end{equation}
For example, for $q=1.1$, it demands that $p_i>>4.54*10^{-5}$ and
for $q=0.7$, that $p_i>>0.036$. Most of the time, probabilities
are not so small and the estimated value of $q$ is very close to
1. So Eq. (\ref{22**}) is satisfied. The sum of entropies of $n$
identical subsystems can be written as
\begin{equation}\label{23}
\sum_{r=1}^nS_q^{A_r}=nS_q^{A_1}=\frac{n}{q-1}(1-\sum_ip_i^q)=
-n\sum_ip_i\ln{p_i}-\frac{n}{2}(q-1)\sum_ip_i{(\ln{p_i})}^2-\ldots~,
\end{equation}
and for the entropy of the composed system, we have
\begin{equation}\label{24}
S_{q'}^{A_1+A_2+\ldots+A_n}=\frac{1}{q'-1}(1-\sum_{i_1i_2...i_N}p_{i_1i_2...i_N}^{q'})~,
\end{equation}
where $p_{i_1i_2...i_N}$s are probabilities related to the
composed system. Because $q$ and $q'$ are close to 1, the
probabilities in different subsystems are nearly independent and,
so
\begin{equation}\label{25}
\sum_{i_1i_2...i_N}p_{i_1i_2...i_N}^{q'}=\sum_{i_1i_2...i_N}p_{i_1}^{q'}p_{i_2}^{q'}\ldots{p_{i_n}^{q'}}
={(\sum_{i}p_{i}^{q'})}^n~.
\end{equation}
Using Eq. (\ref{22}), we have
\begin{equation}\label{26}
{(\sum_{i}p_{i}^{q'})}^n=1+n(q'-1)\sum_ip_i\ln{p_i}+\frac{n}{2}{(q'-1)}^2\sum_ip_i{(\ln{p_i})}^2+
\frac{n(n-1)}{2}{(q'-1)}^2{(\sum_ip_i\ln{p_i})}^2+\ldots~,
\end{equation}
where $|n(q'-1)\ln{p_i}|<<1$, so that the higher-order terms are
very small and negligible. Hence, the entropy of the composed
system is obtained as follows:
\begin{equation}\label{27}
S_{q'}^{A_1+A_2+\ldots+A_n}=-n\sum_ip_i\ln{p_i}-\frac{n}{2}{(q'-1)}\sum_ip_i{(\ln{p_i})}^2-
\frac{n(n-1)}{2}{(q'-1)}{(\sum_ip_i\ln{p_i})}^2+\ldots~.
\end{equation}
Quasi-additivity, thus, implies a relation between $q$ and $q'$,
namely,
\begin{equation}\label{28}
\frac{q'-1}{q-1}=\frac{\sum_ip_i{(\ln{p_i})}^2}{\sum_ip_i{(\ln{p_i})}^2+(n-1){(\sum_ip_i\ln{p_i})}^2}~,
\end{equation}
which can be rewritten in the following simplified form
\begin{equation}\label{29}
\frac{q'-1}{q-1}=\frac{\langle{B_i^2}\rangle}{\langle{B_i^2}\rangle+(n-1){\langle{B_i}\rangle}^2}~.
\end{equation}
For $n=2$, Eq. (\ref{14}) is recovered. The R.H.S. of Eq.
~(\ref{29}) is positive and less than 1, so $\frac{q'-1}{q-1}$ is
also positive and we can write
\begin{equation}\label{29*}
0<|\frac{q'-1}{q-1}|=\frac{\langle{B_i^2}\rangle}{\langle{B_i^2}\rangle+(n-1){\langle{B_i}\rangle}^2}<1,
\hspace{.5cm}n>1~.
\end{equation}
Eq.~(\ref{29*}) shows the dependence of $q'$ on $n$. It also says
that
\begin{equation}\label{210}
|q'-1|<|q-1|\hspace{.2cm}\Rightarrow\hspace{.2cm}\left \{
\begin{array}{ll}
1<q'<q & {\rm if}\ q>1\\
q<q'<1 & {\rm if}\ q<1
\end{array} \right.
\end{equation}
If we assume that $n$ increases by increasing scale $r$, it is
clear from Eq.~(\ref{29*}) that for $q>1$ $(q<1)$, $q'(r)$ is a
strictly monotonously decreasing (increasing) function of scale
$r$, so that by increasing system size, $q'$ approaches 1. From
Eq.~(\ref{12}), it is clear that the deviation of parameter $q$
from 1 can be interpreted as the fluctuations in temperature (or
energy dissipation). Hence, approaching of $q'$ to 1 by increasing
n, indicates that fluctuations become negligible when system size
increases. Assuming that $n$ is proportional to volume
($V\propto{r^3}$), in the limit $n\rightarrow\infty$, it is clear
from Eq.~(\ref{29}) that
\begin{equation}\label{211}
 |q'-1|\propto\frac{1}{n}\propto\frac{1}{V}~.
\end{equation}
The dependence of $q$ on the size of the system has been studied
in the hadronization process \cite{17}, where $q-1$ is introduced
as a measure of total heat capacity of the hadronizing system
\cite{18,19}: $q-1=\frac{1}{C}$, and then the assumption of
$C\propto{V}$ results in: $q-1=\frac{1}{V}$, where $q$ is $q'$ in
our discussion. However, it should be noted that the higher-order
terms in (\ref{26}) are negligible, only if for each $n$ we have
\begin{equation}\label{212}
|n(q'-1)\ln{p_i}|<<1\hspace{.2cm},\hspace{.5cm}\forall{p_i}
\end{equation}
and that this condition should also be confirmed also in the limit
$n\rightarrow\infty$. By noting that
$\frac{\langle{B_i^2}\rangle-{\langle{B_i}\rangle}^2}{{\langle{B_i}\rangle}^2}$
is positive, one can change equality in Eq.~(\ref{29}) to the
following inequalities:
\begin{equation}\label{213}
|n(q'-1)|\leq|q-1|\frac{\langle{B_i^2}\rangle}{{\langle{B_i}\rangle}^2}\Rightarrow
|n(q'-1)\ln{p_i}|\leq|q-1|\frac{\langle{B_i^2}\rangle}{{\langle{B_i}\rangle}^2}|\ln{p_i}|~.
\end{equation}
Hence, condition~(\ref{212}) is confirmed if
\begin{equation}\label{214}
A=|q-1|\frac{\sum_ip_i{(\ln{p_i})}^2}{{(\sum_ip_i\ln{p_i})}^2}|\ln{p_i}|<<1\hspace{.2cm},\hspace{.5cm}\forall{p_i}~.
\end{equation}
We use that condition to find permissible regions for binary
subsystems, which have two possible states with the probabilities
$p$ and $1-p$. The regions have been shown in Fig.~(\ref{**}) for
$A=0.4$ and $A=0.2$. These values of $A$ give a nearly good
approximation for the expansion~(\ref{26}), because we hold the
terms which contain $A^2$.\\

In this section, we studied quasi-additivity relation for n
independent identical subsystems. Our expansion makes sense only
when $q$ is close to 1. It should be noted that 1 is the value of
$q$ for which the Tsallis entropy becomes additive in independent
subsystems. At this point, the question arise as to what we can
say about correlated subsystems if we want to use
quasi-additivity relation for them. The answer is provided in the
following section.

\section{Quasi-additivity property for correlated identical
subsystems} As implied, if subsystems are independent, then the
Tsallis entropy of a composite system becomes additive for
$q\rightarrow1$ (Boltzmann-Gibbs entropy) and for other values of
$q$, the Tsallis entropy is not additive (\ref{15}). However, when
there are some correlations between subsystems, the Tsallis
entropy may be additive for an appropriate $q\neq1$. In this
section, we consider $n$ equal subsystems, which are specially
correlated. So additivity condition is established for $q^*\neq1$:
\begin{equation}\label{31}
\sum_{r=1}^nS_{q^*}^{A_r}=S_{{q^*}}^{A_1+A_2+\ldots+A_n}~.
\end{equation}
It may be interesting to use the quasi-additivity property (Eq.
(\ref{21})) for correlated subsystems. Similar to the case of
independent subsystems where $q$ and $q'$ are assumed to be close
to 1, namely, additive case, here also we assume that $q$ and $q'$
are close to the additive case $q^*$. Supposing that
$q=q^*+\delta$, one obtains
\begin{equation}\label{32}
\sum_ip_i^q=\sum_ip_i^{q^*+\delta}=\sum_i{p_i}^{q^*}\exp^{\delta\ln{p_i}}=
\sum_ip_i^{q^*}+\delta\sum_ip_i^{q^*}\ln{p_i}+\frac{1}{2}{\delta}^2\sum_ip_i^{q^*}{(\ln{p_i})}^2+\ldots~,
\end{equation}
where only if $|\delta~{\ln{p_i}}|<<1$, it is possible to neglect
the  higher-order terms in the expansion. We have a coefficient
$\frac{1}{q-1}$ in the entropy and are interested in expanding it
to powers of $\delta$
\begin{eqnarray}
&&\frac{1}{q-1}=\frac{1}{q^*-1+\delta}=\frac{q^*-1-\delta}{{(q^*-1)}^2-\delta^2}=\frac{q^*-1-\delta}{{(q^*-1)}^2}
(1+\frac{\delta^2}{{(q^*-1)}^2}+\ldots)\nn\\
&&\hspace{1.1cm}=\frac{1}{q^*-1}-\frac{\delta}{{(q^*-1)}^2}+\frac{\delta^2}{{(q^*-1)}^3}+\ldots~.\label{33}
\end{eqnarray}
So the entropy $S_q^{A_r}$ can be written as
\begin{eqnarray}
&&S_q^{A_r}=S_q^{A_1}=\frac{1-\sum_ip_i^q}{q-1}\label{34}\\
&&\hspace{.75cm}=(1-\sum_ip_i^{q^*}-\delta\sum_ip_i^{q^*}\ln{p_i}-\frac{1}{2}{\delta}^2\sum_ip_i^{q^*}{(\ln{p_i})}^2+\ldots)
(\frac{1}{q^*-1}-\frac{\delta}{{(q^*-1)}^2}+\frac{\delta^2}{{(q^*-1)}^3}+\ldots),\nn
\end{eqnarray}
and to second-order of $\delta$ we obtain
\begin{equation}\label{35}
S_q^{A_r}=\frac{1-\sum_ip_i^{q^*}}{{q^*}-1}-\delta\Big(\frac{1-\sum_ip_i^{q^*}}{{(q^*-1)}^2}
+\frac{\sum_ip_i^{q^*}\ln{p_i}}{{q^*}-1}\Big)+\delta^2\Big(\frac{1-\sum_ip_i^{q^*}}{{(q^*-1)}^3}+
\frac{\sum_ip_i^{q^*}\ln{p_i}}{{(q^*-1)}^2}-\frac{1}{2}\frac{\sum_ip_i^{q^*}{(\ln{p_i})}^2}{{q^*}-1}\Big)+\ldots~.
\end{equation}
Similarly for the composed system, one can write
\begin{eqnarray}\label{36}
&&S_{q'}^{A_1+A_2+\ldots+A_n}=\frac{1-\sum{p}_{i_1i_2...i_n}^{q^*}}{{q^*}-1}
-\varepsilon\Big(\frac{1-\sum{p}_{i_1i_2...i_n}^{q^*}}{{(q^*-1)}^2}
+\frac{\sum{p}_{i_1i_2...i_n}^{q^*}\ln{p_{i_1i_2...i_n}}}{{q^*}-1}\Big)\nn\\
&&\hspace{1.5cm}+\varepsilon^2\Big(\frac{1-\sum{p}_{i_1i_2...i_n}^{q^*}}{{(q^*-1)}^3}+
\frac{\sum{p}_{i_1i_2...i_n}^{q^*}\ln{p_{i_1i_2...i_n}}}{{(q^*-1)}^2}-
\frac{1}{2}\frac{\sum{p}_{i_1i_2...i_n}^{q^*}{(\ln{p_{i_1i_2...i_n}})}^2}{{q^*}-1}\Big)+\ldots~,
\end{eqnarray}
where $\varepsilon=q'-q^*$, the sum is over all $i_r$ and
$P_{i_1i_2...i_n}$s satisfy extensivity condition (Eq.
(\ref{31})). From the above relations, it is clear that the
entropies $S_q^{A_r}$ and $S_{{q}}^{A_1+\ldots+A_n}$ in the first
approximation are equal to $S_{q^*}^{A_r}$ and
$S_{{q^*}}^{A_1+\ldots+A_n}$, respectively, where $q^*$
corresponds to the extensive case. Using quasi-additivity
property, Eq. (\ref{21}) in conjunction with Eq. (\ref{31}) yields
\begin{eqnarray}\label{37}
&&n\delta\Big(\frac{1-\sum_ip_i^{q^*}}{{(q^*-1)}^2}
+\frac{\sum_ip_i^{q^*}\ln{p_i}}{{q^*}-1}\Big)
-n\delta^2\Big(\frac{1-\sum_ip_i^{q^*}}{{(q^*-1)}^3}+
\frac{\sum_ip_i^{q^*}\ln{p_i}}{{(q^*-1)}^2}-\frac{1}{2}\frac{\sum_ip_i^{q^*}{(\ln{p_i})}^2}{{q^*}-1}\Big)\nn\\
&&=\varepsilon\Big(\frac{1-\sum{p}_{i_1i_2...i_n}^{q^*}}{{(q^*-1)}^2}
+\frac{\sum{p}_{i_1i_2...i_n}^{q^*}\ln{p_{i_1i_2...i_n}}}{{q^*}-1}\Big)\nn\\
&&-\varepsilon^2\Big(\frac{1-\sum{p}_{i_1i_2...i_n}^{q^*}}{{(q^*-1)}^3}+
\frac{\sum{p}_{i_1i_2...i_n}^{q^*}\ln{p_{i_1i_2...i_n}}}{{(q^*-1)}^2}-
\frac{1}{2}\frac{\sum{p}_{i_1i_2...i_n}^{q^*}{(\ln{p_{i_1i_2...i_n}})}^2}{{q^*}-1}\Big)~.
\end{eqnarray}
The above relation can be rewritten as
\begin{equation}\label{38}
n\delta{A}-n\delta^2B=\varepsilon{C}-\varepsilon^2D~,
\end{equation}
where
\begin{eqnarray}
&&A=\frac{1-\sum_ip_i^{q^*}}{{(q^*-1)}^2}
+\frac{\sum_ip_i^{q^*}\ln{p_i}}{{q^*}-1}~,\label{39}\\
&&B=\frac{A}{q^*-1}-\frac{1}{2}\frac{\sum_ip_i^{q^*}{(\ln{p_i})}^2}{{q^*}-1}~,\label{310}\\
&&C=\frac{1-\sum{p}_{i_1i_2...i_n}^{q^*}}{{(q^*-1)}^2}
+\frac{\sum{p}_{i_1i_2...i_n}^{q^*}\ln{p_{i_1i_2...i_n}}}{{q^*}-1}~,\label{311}\\
&&D=\frac{C}{q^*-1}-\frac{1}{2}\frac{\sum{p}_{i_1i_2...i_n}^{q^*}{(\ln{p_{i_1i_2...i_n}})}^2}{{q^*}-1}~.\label{312}
\end{eqnarray}
By defining
\begin{equation}\label{313}
\gamma\equiv\frac{\varepsilon}{\delta}\equiv\frac{q'-q^*}{q-q^*}~,
\end{equation}
Eq. (\ref{38}) can be rewritten as
\begin{equation}\label{314}
D\gamma^2-\frac{C}{\delta}\gamma+\frac{nA}{\delta}-nB=0~,
\end{equation}
and its solution is obtained as follows
\begin{equation}\label{315}
\gamma=\frac{1}{2D}\Big(\frac{C}{\delta}\pm\sqrt{\frac{C^2}{\delta^2}-4Dn(\frac{A}{\delta}-B)}\Big)
=\frac{1}{2D}\Big(\frac{C}{\delta}\pm|\frac{C}{\delta}|\sqrt{1-\frac{4Dn}{C^2}(A\delta-B\delta^2)}\Big)~.
\end{equation}
It is observed that  there are two solutions which contain  the
term $\frac{C}{2D\delta}$. Because $\delta$ is small,
$\frac{C}{2D\delta}$ becomes very large. But we are interested in
finding finite solutions. If the coefficients of $\delta$ and
$\delta^2$ in the radical are finite, then Eq. (\ref{315}) can be
rewritten as
\begin{equation}\label{316}
\gamma=\frac{C}{2D\delta}\pm\frac{|C|}{2D|\delta|}\Big(1-\frac{2DAn}{C^2}\delta
+(\frac{2DBn}{C^2}-\frac{2D^2A^2n^2}{C^4})\delta^2\Big)~.
\end{equation}
We can choose `+' or `$-$' signs from Eq. (\ref{316}) (depending
on the sign of $|\frac{C}{\delta}|$) so that the terms
proportional to $\delta^{-1}$ are eliminated and the solution is
finite. It should be noted that when $q^*\rightarrow1$ or in cases
where probabilities are very small so that their logarithms
become infinite, the coefficients $A,B,C$ and $D$ can be infinite
and, hence, Eq. (\ref{316}) will not be valid.

\section{Finding probabilities which satisfy additivity relation}
We assumed that $p_{i_1i_2...i_n}$ s are the probabilities which
make the Tsallis entropy additive, but there is a problem of
finding those probabilities for each $n$. We restrict ourselves
to $n$ equal binary subsystems which have been investigated in
\cite{14} by Tsallis, Gell-Mann and Sato. Because our subsystems
are binary, we have two possible microstates for each subsystem.
Namely, for $n=1$, the probabilities of microstates are $p_1=p$
and $p_2=1-p$. For $n=2$, the probabilities are $p_{11}$,
$p_{12}=p_{21}$ and $p_{22}$ and they satisfy the following
relations
\begin{eqnarray}
&&p_{11}+p_{12}=p_1=p\label{41}~,\\
&&p_{21}+p_{22}=p_2=1-p~.\label{42}
\end{eqnarray}
Eqs. (\ref{41}) and (\ref{42}) imply  that for two subsystems $A$
and $B$, the sum of probabilities of the composed system $A+B$
over microstates of $B$ with a specified microstate of $A$
results in the probability of that microstate of $A$, which is
reasonable. Adding Eq. (\ref{41}) to (\ref{42}) yields
\begin{equation}\label{43}
p_{11}+2p_{12}+p_{22}=1~.
\end{equation}
In the case of independent subsystems, $p_{11}=p^2,
p_{12}=p(1-p)$ and $p_{22}=(1-p)^2$, so  Eq. (\ref{43}) becomes
trivial. Similarly, for three equal binary subsystems, we can
write
\begin{eqnarray}
&&p_{111}+p_{112}=p_{11}~,\label{44}\\
&&p_{121}+p_{122}=p_{12}~,\label{45}\\
&&p_{221}+p_{222}=p_{22}~,\label{46}
\end{eqnarray}
and with the help of Eqs. (\ref{43}) to (\ref{46}), we obtain
\begin{equation}\label{47}
p_{111}+3p_{112}+3p_{122}+p_{222}=1~,
\end{equation}
where $p_{112}=p_{121}=p_{211}$ and $p_{122}=p_{212}=p_{221}$.
Eqs. (\ref{41}), (\ref{42}) and (\ref{44}) to (\ref{46}) are
referred to as ``scale-invariance'' ( or scale-freedom) relations
which can be generalized to n subsystems. They are called
`Leibnitz rule' in another notation \cite{14} which relates the
probabilities of $n$ systems to the probabilities of $n-1$
systems. Eq. (\ref{47}) can be generalized to $n$ subsystems
using `Pascal~triangle'. However, in every stage, the number of
equations are 1 less than the number of unknown probabilities. So
it is not possible to exactly find the probabilities. For
example, for $n=2$, we have two Eqs. (\ref{41}) and (\ref{42}) and
three unknown probabilities $p_{11}$, $p_{12}$ and $p_{22}$.
Another equation that can be added to find our probabilities is
additivity condition for entropy. In this way, for two equal
binary subsystems, it can be written as
\begin{equation}\label{48}
S_{q^*}^{A+B}=2S_{q^*}^{A}\hspace{.5cm}\Rightarrow\hspace{.5cm}
\frac{1-p_{11}^{q^*}-2p_{12}^{q^*}-p_{22}^{q^*}}{q^*-1}=2\hspace{.1cm}\frac{1-p^{q^*}-(1-p)^{q^*}}{q^*-1}~,
\end{equation}
and with the help of scale-invariance conditions, we have
\begin{equation}\label{49}
\frac{1-p_{11}^{q^*}-2(p-p_{11})^{q^*}-(1-2p+p_{11})^{q^*}}{q^*-1}=2\hspace{.1cm}\frac{1-p^{q^*}-(1-p)^{q^*}}{q^*-1}~.
\end{equation}
So $p_{11}$, $p_{12}$ and $p_{22}$ are obtained to show that
additivity can exist for correlated subsystems with $q^*\neq1$.
After finding the probabilities of two composed systems, it is
possible to obtain the probabilities of three composed systems
using scale-invariance and additivity conditions. These processes
may be continued untill all the probabilities which make the
entropy additive for each $n$ are found.

\section{Numerical solution of quasi-additivity relation for two
correlated equal binary subsystems}
\subsection{Exact solution:}
Similar to the additive case, quasi-additivity relation for two
correlated identical binary subsystems can be written as
\begin{equation}\label{51}
S_{q'}^{A+B}=2S_{q}^{A}\hspace{.5cm}\Rightarrow\hspace{.5cm}
\frac{1-p_{11}^{q'}-2p_{12}^{q'}-p_{22}^{q'}}{q'-1}=2\hspace{.1cm}\frac{1-p^{q}-(1-p)^{q}}{q-1}~,\end{equation}
where $p_{11}$, $p_{12}=p-p_{11}$ and $p_{22}=1-2p+p_{11}$ are
the probabilities related to the additive case and can be obtained
from Eq. (\ref{49}). For a specified $p$ and $q$ ($q$ is assumed
to be close to $q^*$ as in the case of independent subsystems
where $q$ is close to 1), it is possible to find $q'$ from Eq.
(\ref{51}) and compare it with $q$. Also $\gamma$ is obtained from
Eq. (\ref{313}), which we designate as $\gamma_e$ to represent the
word `exact'.
\subsection{Approximate solution:}
It is not always possible to find the solution of quasi-additivity
relation for $n$ correlated subsystems exactly.  One may,
therefore, use the approximate solution given in (\ref{315}). If
we are now interested in finding $q'$ or $\gamma$ for two equal
binary correlated subsystems with specified $p,q$ and $q^*$ using
our approximate relation, then for $A,B,C$ and $D$ from Eqs.
(\ref{39}) to (\ref{312}) we obtain
\begin{eqnarray}
&&A=\frac{1-p^{q^*}-(1-p)^{q^*}}{{(q^*-1)}^2}
+\frac{p^{q^*}\ln{p}+(1-p)^{q^*}\ln{(1-p)}}{{q^*}-1}~,\label{52}\\
&&B=\frac{A}{q^*-1}-\frac{1}{2}\frac{p^{q^*}{(\ln{p})}^2+(1-p)^{q^*}{(\ln{(1-p)})}^2}{{q^*}-1}~,\label{53}\\
&&C=\frac{1-{p}_{11}^{q^*}-2{p}_{12}^{q^*}-{p}_{22}^{q^*}}{{(q^*-1)}^2}
+\frac{{p}_{11}^{q^*}\ln{p}_{11}+2{p}_{12}^{q^*}\ln{p}_{12}+{p}_{22}^{q^*}\ln{p}_{22}}{{q^*}-1}~,\label{54}\\
&&D=\frac{C}{q^*-1}-\frac{1}{2}\frac{{p}_{11}^{q^*}{(\ln{p_{11}})}^2+2{p}_{12}^{q^*}{(\ln{p_{12}})}^2+
{p}_{22}^{q^*}{(\ln{p_{22}})}^2}{{q^*}-1}~.\label{55}
\end{eqnarray}
So from Eq. (\ref{315}), $\gamma$ is obtained and designated as
$\gamma_a$, where $a$ represents `approximate'. We can, then,
obtain $q'$ from Eq. (\ref{313}).
\subsection{Results}

The results are given in Tables 1 and 2 for $q^*=0.3$ and
$q^*=0.5$ and for a given $q$ close to $q^*$ as $p$ changes from
0.1 to 0.9 in columns. $p_{11}, p_{12}$ and $p_{22}$ (the
probabilities related to the additive case) are given in the three
first columns. The results of the exact solution of
quasi-additivity equation (section~  5.1) and also approximate
solution (section~ 5.2) are shown in the following columns. It is
seen that in both approximate and exact solutions, $\gamma$ is
positive and less than 1, namely, $|{q'-q^*}|<|{q-q^*}|$ which
means $q$ approaches $q^*$ by increasing system size, or
 alternatively, spatial scales which are similar to independent
subsystems where $q'$ approaches 1 by increasing system size. It
is an interesting result which means that by increasing spatial
scales, fluctuations become negligible.

\section{Conclusion}
In this paper, Beck's concept of quasi-additivity was used for $n$
independent subsystems and the relation between $q$ and $q'$ was
found for $q$ and $q'$ to be close to 1 (Eq. (\ref{29})). It
strictly shows that by increasing $n$, or system size, $q'$
approaches 1 so that $q'\rightarrow1$ when $n\rightarrow\infty$.
The deviation of the parameter $q$ from 1 in non-equilibrium
systems describes the fluctuations in temperature (or energy
dissipation)(Eq. (\ref{12})). Approaching of $q'$ to 1 by
increasing $n$ shows how fluctuations become negligible when
system size increases. We have also used the quasi-additivity
relation for correlated subsystems where it is assumed that $q$
and $q'$ are close to $q^*$ (specific value of $q$, which makes
Tsallis entropy additive). It has been found that for the case of
correlated subsystems, $q$ also approaches $q^*$ which depicts a
 behaviour similar to the case of independent subsystems. The results
 for the two correlated equal binary subsystems have been given in
 the Tables.

\newpage
\begin{table}
\centering {\footnotesize\renewcommand{\baselinestretch}{0.4}
\begin{tabular}{||p{1.6cm}p{1.6cm}p{1.6cm}p{1.6cm}p{1.6cm}p{1.6cm}||}\cline{1-6}
  ${q=0.35}$&${q^*=0.3}$ & & & & \\
  ${q=0.25}$& $p_{11}$ & $p_{12}$ & $p_{22}$ & $\gamma_e$ &
   $\gamma_a$  \\ \hline
  $p=0.1$ & 0.09  & 0.01  & 0.89  &
$\_\_\_\_\_ $  & 0.52  \\
  & 0.09  & 0.009  & 0.89  & 0.52 & 0.52 \\
\cline{1-6} \hline
  $p=0.3$ & 0.28 & 0.02  & 0.68  & 0.32 & 0.32 \\
  & 0.28 & 0.02  & 0.68  & 0.32  &  0.32 \\
\cline{1-6} \hline
  $p=0.5$ & 0.02 & 0.48  & 0.02  & 0.26 & 0.26 \\
  & 0.02 & 0.48  & 0.02  & 0.25  & 0.25 \\
\cline{1-6}\hline
  $p=0.7$ & 0.68 & 0.02  & 0.28  & 0.32 & 0.32 \\
  & 0.68 & 0.02  & 0.28  & 0.32  & 0.32 \\
\cline{1-6} \hline
  $p=0.9$ & 0.89 & 0.01  & 0.09  &
$\_\_\_\_\_ $ & 0.52\\
  & 0.89 & 0.01  & 0.09  & 0.52  &  0.52 \\
\cline{1-6} \hline \hline\cline{1-6}
\end{tabular}}
\caption{\footnotesize $\gamma_e$ and $\gamma_a$ are given for
$q^*$=0.3. The  upper lines in the Tables correspond to $q=0.35$
and the lower ones to $q=0.25$. Results are rounded to two digits
after decimal point. As shown, $\gamma_e$ and $\gamma_a$ are less
than 1. }\label{1}
\end{table}

\begin{table}
\centering {\footnotesize\renewcommand{\baselinestretch}{0.4}
\begin{tabular}{||p{1.6cm}p{1.6cm}p{1.6cm}p{1.6cm}p{1.6cm}p{1.6cm}||}\cline{1-6}
  ${q=0.55}$ &${q^*=0.5}$  &  &  &  &  \\
  ${q=0.45}$ & $p_{11}$ & $p_{12}$ & $p_{22}$ & $\gamma_e$ & $\gamma_a$ \\ \hline
  $p=0.1$ & 0.07 & 0.03  & 0.87  & 0.62 & 0.62  \\
  & 0.07 & 0.03  & 0.87  & 0.63  &  0.63 \\
\cline{1-6} \hline
  $p=0.3$ & 0.24 & 0.06  & 0.64  & 0.43 & 0.43 \\
  & 0.24 & 0.06  & 0.64  & 0.43  & 0.43 \\
\cline{1-6} \hline
  $p=0.5$ & 0.44 & 0.06  & 0.44  & 0.37 & 0.37 \\
  & 0.44 & 0.06  & 0.44  & 0.36  &  0.36 \\
\cline{1-6} \hline
  $p=0.7$ & 0.64 & 0.06  & 0.24  & 0.43 & 0.43 \\
  & 0.64 & 0.06  & 0.24  & 0.43  &  0.43 \\
\cline{1-6} \hline
  $p=0.9$ & 0.87 & 0.03  & 0.07  & 0.62 & 0.62 \\
  & 0.87 & 0.03  & 0.07  & 0.63  & 0.63 \\
\cline{1-6} \hline \hline\cline{1-6}
\end{tabular}}
\caption{\footnotesize $\gamma_e$ and $\gamma_a$ are given for
$q^*$=0.5. The upper lines in the Tables correspond to $q=0.55$
and the lower ones to $q=0.45$. Results are rounded to two digits
after decimal point. As shown, $\gamma_e$ and $\gamma_a$ are less
than 1.}\label{2}
\end{table}

\end{document}